%% file: main.tex
\documentclass[sigconf, nonacm]{acmart}

\newcommand\vldbdoi{XX.XX/XXX.XX}

\newcommand\vldbavailabilityurl{https://github.com/huwf1220/cfp_parallel.git}
\newcommand\vldbpagestyle{plain} 

\usepackage{balance}

\usepackage{amsmath,amssymb,amsfonts}
\usepackage{graphicx}
\usepackage{textcomp}
\usepackage{xcolor}

\usepackage{tikz}
\usepackage{amsmath}

\usepackage{filecontents}
\usepackage{xspace}
\usepackage{algorithmicx}
\usepackage{algpseudocode} 
\usepackage{amsmath}
\usepackage{booktabs}
\usepackage{adjustbox}
\usepackage[linesnumbered, ruled, vlined]{algorithm2e}
\usepackage[normalem]{ulem}
\usepackage{soul}

\SetKwInOut{Input}{Input}
\SetKwInOut{Output}{Output}
\newcommand{\sysname}{{\em CFP}\xspace}
\newcommand{\pb}{ParallelBlock\xspace}

\begin{document}
\title{\sysname: Low-overhead Profiling-based Intra-operator Parallelism Generation by Preserving Communication-Free Structures}

\author{Weifang Hu$^{1}$, Xuanhua Shi$^{1}$,  Yunkai Zhang$^{1}$, Chang Wu$^{1}$, Xuan Peng$^{1}$, Jiaqi Zhai$^{1}$, Hai Jin$^{1}$, Xuehai Qian$^{2}$, Jingling Xue$^{3}$, Yongluan Zhou$^{4}$}

\affiliation{
  \institution{$^{1}$Huazhong University of Science and Technology \quad
               $^{2}$Tsinghua University \quad
               $^{3}$UNSW Sydney \quad
               $^{4}$University of Copenhagen}
}
\email{{huwf, xhshi, m202373843, wuchang, piecesix, jqzhai, hjin}@hust.edu.cn}
\email{xuehaiq@tsinghua.edu.cn, jingling@cse.unsw.edu.au, zhou@di.ku.dk}

\begin{abstract}
Optimizing the parallel training of large models requires exploring intra-operator parallelism plans for a computation graph that typically contains tens of thousands of primitive operators. While the optimization of parallel data processing graphs has been extensively researched in database systems, the vast search space makes it challenging to apply traditional database query optimization methods and algorithms. This paper introduces \sysname, an optimization system for intra-operator parallelism that significantly reduces the complexity of searching for parallelism plans by leveraging two structural patterns found in large models. First, we identify parallel-preserving subgraphs, which ensure that the optimal global plan assigns the same parallel strategy to all operators within the subgraph. This approach allows us to avoid enumerating all possible combinations of parallel strategies for these operators. Second, we recognize repetitive subgraph patterns within the large computational graph, enabling us to profile a moderate number of representative subgraphs and accurately estimate the cost of parallelism plans with low overhead. With the significantly reduced search space, we can employ dynamic programming to search for the optimized parallelism plan. In our experiments, we demonstrate that \sysname achieves significant speedups 
compared to the state-of-the-art framework for large models like GPT and LLAMA.
\end{abstract}

\maketitle

\pagestyle{\vldbpagestyle}


\begingroup
\let\clearpage\relax
\include{content/Introduction}
\include{content/Background}

\include{content/Method/ParallelBlock}

\include{content/Method/Profile}

\include{content/Evaluation/Setup}

\include{content/Evaluation/E2EPerf}

\include{content/Evaluation/Analysis}

\include{content/Evaluation/MemConstraint}

\include{content/Evaluation/CompileOverhead}

\include{content/Evaluation/CaseStudy}

\include{content/RelatedWorks}

\include{content/Limitation}
\endgroup

\balance
\bibliographystyle{ACM-Reference-Format}
\bibliography{ref}

\end{document}

%% file: content/Introduction.tex
\section{Introduction}\label{intro}

Recent advances in large models have propelled remarkable progress in diverse fields~\cite{yang2023harnessing, kirillov2023segment, RN245}. 
These models, characterized by their extensive parameter sizes and computational requirements, necessitate the collaborative processing power of hundreds or even thousands of GPUs over extended periods. 
Recent efforts have extended the training of large models to over 10K GPUs~\cite{megascale, llama3}. 
This has made optimizing parallelism in large models a critical challenge. 

Motivated by the massive data scale and computational demands in large model training, modern training systems have widely adopted multi-level parallelism configurations: from coarse-grained or inter-operator parallelism~\cite{narayanan2019pipedream, huang2019gpipe, AdaPipe} where the model is partitioned into segments handled by different GPUs or GPU groups, to fine-grained or intra-operator parallelism where the computation within each operator is split across multiple GPUs.
\textcolor{black}{Intra-operator parallelism can be applied to the whole model or orthogonally to model segments in inter-operator parallelism, e.g., stages in pipeline parallelism.}
This paper focuses on automatically searching for
efficient intra-operator parallelism plans,
which is crucial for the end-to-end performance of large model training~\cite{auto-parallel-survey}.

\textcolor{black}{
Intra-operator parallelism involves partitioning the computation of each operator in the model across different devices and running them in a Single Program Multiple Data (SPMD) form to improve training throughput and reduce memory usage~\cite{tofu2019, hap_heter_gpu_spmd_eurosys24}.
For example, a two-dimensional matrix multiplication can be partitioned by rows, columns, or reduction dimension, which exemplify an operator’s parallelization strategies, i.e., how a single operator’s computation is divided among multiple GPUs.
Each operator is assigned a parallelism strategy, and the combination of these individual strategies forms the overall parallelism plan of the entire model. 
Different intra-operator parallelism plans expose different communication and computation patterns, which can lead to significant differences in parallelism efficiency~\cite{xu2021gspmd}. 
}

Although manually designed intra-operator parallelism plan templates~\cite{li2020pytorch,megatronlm21sc,fsdp_pytorch}, such as tensor parallelism, data parallelism, or fully sharded data parallelism (FSDP), have achieved promising results, they require substantial development and tuning efforts~\cite {lepikhin2020gshardscalinggiantmodels}. 
A more general approach is to transform the model into a logical computation graph composed of primitive operators (e.g., matrix multiplication and arithmetic operations), and then automatically assign parallelism strategies to each operator to form a logical parallelism plan. Recent works have introduced many tensor partition abstractions for fine-grained operators~\cite{xu2021gspmd, santhanam2021distir, primpar_eurosys24}, thereby forming a comprehensive intra-operator parallel search
spaces~\cite{lu2017flexflow, zheng2022alpa, tofu2019}.

To quickly estimate the performance of different parallelism plans in the search space, most existing approaches adopt symbolic cost models based on static information. These models mainly consider the communication volume and the number of floating-point operations (FLOPs) required by each plan~\cite{zheng2022alpa, tofu2019, accpar, pase, cai2021tensoropt, D_rec}.
Some other works explore machine learning-based models~\cite{icml17_rl_cost_model, gdp_rl_cost_model, lstm_cost_model, wang2020automap, lu2017flexflow} to capture the complex relationships among various factors and better model the impact of parallelism plans. 

These static models, whether symbolic or machine learning-based, primarily estimate the cost of parallelism plans of the logical computation graph. However, the logical computation graph must be compiled and optimized into a physical graph for execution. Consequently, the estimated cost often fails to align with the actual execution cost. For example, a compiler may automatically fuse multiple operators to optimize communication efficiency~\cite{nsdi22_optimize_comm_for_dp, icdcs23_opt_all_reduce}, resulting in a physical computation graph whose cost sufficiently deviates from that of the logical one.  
Furthermore, many factors of the runtime environment can increase the performance uncertainty. For instance, the actual performance of the underlying communication libraries is highly sensitive to the runtime tensor size~\cite{asplos22_synthesis_comm_kernel, nsdi23_taccl_comm_library}, partly due to their hardware-dependent optimization. These unpredictable performance characteristics severely undermine the reliability of static models. Incorporating all these factors into a static model would overly complicate the model, making it sensitive to subtle variations in the target runtime environment.

To tackle these challenges, we can learn from query optimizers in database systems, which focus on estimating the cost of physical execution plans rather than logical plans. Additionally, to improve the robustness of the optimizer in uncertain execution environments, we draw inspiration from sampling-based query optimization~\cite{sampling_based_query}, which estimates the cost of query plans by executing queries on samples of data. More specifically, we propose a profiling-based method to robustly estimate the cost of the physical execution plan without requiring a priori knowledge of the actual compiler or execution environment. The idea is that the cost of each parallelism plan is estimated by compiling and running it on sampled data. However, simply borrowing the idea from database query optimizers will not work because the number of operators in a large model (e.g., on the order of tens of thousands) is typically much larger than that of a database query plan. Therefore, it calls for a much more efficient way to search and profile the parallelism plans of large models. Note that optimizing the parallelization of a computation graph has been proved to be NP-hard~\cite{partition_problem_nphard}. 

While the large number of operators in a computation graph leads to exponential growth in the parallelism plan space, we observe two key structural features that can be used to simplify the exponential plan space into a more manageable one, allowing for efficient and robust optimization of parallelism plans.

1) \textbf{Parallelism-preserving subgraphs.} 
A computation graph generally contains many subgraphs, within which all operators can be parallelized using the same strategy to eliminate the need for communication or synchronization between parallel operator instances caused by tensor re-partitioning.   We refer to such subgraphs as {\em parallelism-preserving} subgraphs.
We propose a new coarse-grained structural abstraction, termed \pb.
Each \pb contains a parallelism-preserving subgraph such that the optimal parallel plan would adopt a consistent parallelization strategy for all operators in the \pb.
Identifying parallelism-preserving subgraphs requires analyzing element-level data dependencies between operators, i.e., building the data mapping relations from source to destination at the tensor element level.
Since such dependencies are not directly available from the computation graph, we propose a novel method that constructs element-level dependencies between operators and identifies {\pb}s in a large computation graph efficiently. 
By collapsing all the operators within a \pb into one vertex, we can derive a significantly smaller computation graph. More importantly, our approach ensures that the reduced search space includes the globally optimal plan.

2) \textbf{Repetitive structural patterns.}
A large model typically consists of a limited number of repetitive structural patterns, e.g., identical layers repeated throughout the model. 
By identifying the repetitive structural patterns, we can extract and profile only a small set of representative subgraphs. 
Reusing the profiling results from these subgraphs enables us to estimate the cost of the entire model with significantly lower complexity while maintaining accuracy.
As the compressed graph remains very large (despite being significantly smaller than the original one), identifying the repetitive distinct patterns is computationally expensive. 
To solve this problem, we propose a novel mechanism to generate fingerprints of \pb sequences.
The fingerprint can be used to match different \pb sequences that share consistent sub-parallelism plan spaces and communication behavior.
As a result, our approach converts the parallelism plan space into a much smaller profiling space that scales with the structural diversity of the computation graph rather than the depth of the model.

Finally, we estimate the cost of each segment's parallelism plans using the profiling results and then use a \emph{dynamic programming} algorithm to optimize the plans of each segment, which can then be combined into a global intra-operator parallelism plan with the lowest total cost. 
We implemented the above analysis and optimization in the XLA compiler framework, leveraging its frontend and backend to build an automatic parallelization system, \sysname ({\em C}ommunication-{\em F}ree {\em P}reserve), for large model training. \sysname has been extensively evaluated across four models in various configurations on two target platforms. 
It outperforms state-of-the-art 
frameworks, TensorFlow-Alpa, achieving up to 1.51x, 1.31x, and 3.43x speedup on GPT, LLAMA, and MoE models, respectively. It optimizes parallelism plans for each model in less than 15 minutes.

%% file: content/Background.tex
\vspace{-1ex}
\section{Background and Motivation}\label{background_motivation}

\begin{figure*}
    \centering
    \includegraphics[width=0.98\textwidth]{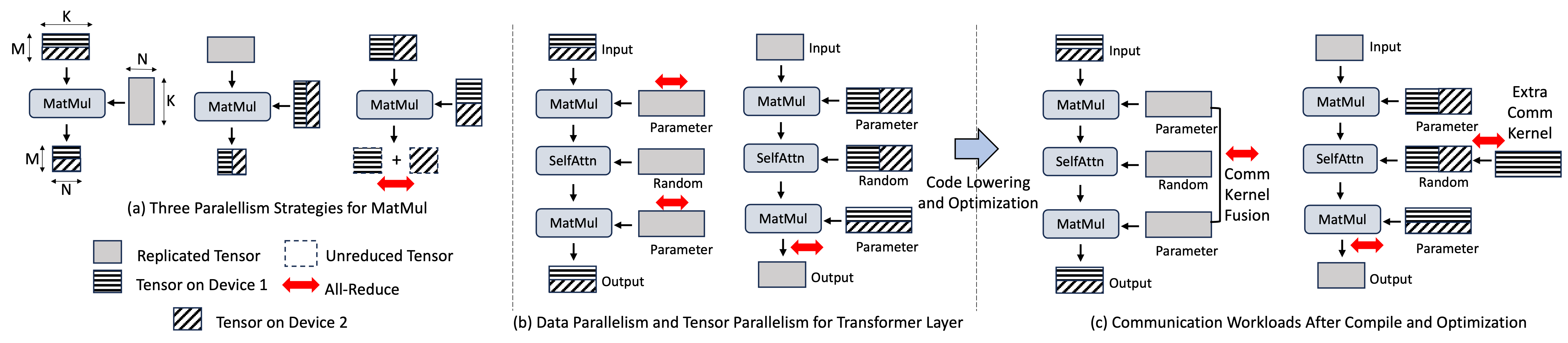}
    \vspace{-3mm}
    \caption{Parallelism strategies for MatMul and two parallelism plans for a Transformer layer. After code lowering and optimizations, there is a noticeable mismatch between the actual communication workloads and the theoretical cost. }
    \vspace{-4mm}
    \label{fig:motivation}
\end{figure*}

\subsection{Parallelism Strategy for Operators}


Model training involves a forward pass, which computes intermediate activations by applying the model's parameters to the input data, and a backward pass, which computes the gradient of the loss with respect to each parameter.
These computations are typically represented by a computation graph, which captures the sequence of operations applied to multi-dimensional tensors.

Operators in the computation graph can be broadly categorized into three types:
(1) Tensor contraction operators, such as matrix multiplication (MatMul) and reduce. These operators are typically compute-intensive and often change the shape of tensors by reducing or expanding specific dimensions.
(2) Element-wise operators, such as addition, subtraction, and multiplication, which apply the same operation independently to each tensor element.
(3) Data manipulation operators, such as reshape and transpose, which change the layout or shape of tensors. 

In intra-operator parallelization, operators are executed on target devices in a SPMD  fashion.
These devices are typically organized into a logical \textit{device mesh}, where they are arranged into a multi-dimensional grid.
For example, 8 GPUs on a node can be organized as a $1 \times 8$ mesh with a single axis, or as a $2 \times 4$ mesh with two axes.
Based on this, for an operator $f: X \rightarrow Y$: $X \in \mathbb{R}^{d_{1} \times \cdots \times d_{k}}$ and $Y \in \mathbb{R}^{d'_{1} \times \cdots \times d'_{l}}$, a \textit{parallelism strategy} specifies how to partition its input tensors over the device mesh and how to distribute the computation accordingly.
Formally, it specifies a subsets of tensor dimension indices, $P_{\text{X}} \subseteq \{1, \dots, k\}$, indicating which dimension of the input tensor are mapped to each axis of the device mesh.
The data along each mapped tensor dimension is then partitioned across the set of devices along the corresponding mesh axis.
As a result, the input and output tensors are divided into \textit{parallel partitions}, i.e., data slices assigned to individual devices for local computation.
If $P_{\text{X}}$ is empty, the input tensor is not partitioned and is instead fully replicated across the devices.

Given the input partition dimensions, the partitioning of an operator's output tensor is constrained by its semantics and the data dependencies between input and output.
Therefore, we refer to the input dimensions in $P_{\text{X}}$ as the operator’s \textit{parallel dimensions}, as they determine how data and computation are partitioned across devices.
When an operator has multiple inputs and outputs, the definition naturally extends by assigning parallel dimensions to each input and output tensor individually.
Moreover, the parallel dimensions across multiple input tensors must be chosen compatibly, so that the operator can be correctly executed on partitioned data without violating operator's semantics.


Taking matrix multiplication in Fig.~\ref{fig:motivation}(a) as an example, the operator 
$f: (A, B) \mapsto C$, where $A \in \mathbb{R}^{M \times K}$, $B \in \mathbb{R}^{K \times N}$, and $C \in \mathbb{R}^{M \times N}$ 
performs a tensor contraction along the shared dimension $K$ and can be parallelized by selecting different parallel dimensions from the input tensors $A$ and $B$.
Specifically, there are three commonly used parallelism strategies:
(1) Row-wise parallelism: choose $P_A = \{1\}$ (the $M$ dimension of $A$) and $P_B = \emptyset$. Each device holds a subset of rows of $A$ and the full $B$, and computes a corresponding row block of $C$.
(2) Column-wise parallelism: choose $P_A = \emptyset$ and $P_B = \{2\}$ (the $N$ dimension of $B$). Each device holds a subset of columns of $B$ and the full $A$, and computes a corresponding column block of $C$.
(3) Split-K parallelism: choose $P_A = \{2\}$ and $P_B = \{1\}$ (the shared $K$ dimension).
Each device holds slices of $A$ and $B$ along the reduction dimension and computes partial results for $C$, which are then aggregated via an inter-device communication (e.g., All-Reduce) to produce the final output.


\subsection{Intra-operator Parallelism Plans}

An \textit{intra-operator parallelism plan} is the global assignment of parallelism strategies to all operators in the computation graph, collectively determining the overall pattern of computation and inter-device communication.
When a tensor produced by one operator is consumed by another, their parallelism strategies may be incompatible, i.e., they partition the shared tensor along different dimensions.
Such mismatches require resharding of the tensor at operator boundaries, introducing inter-device communication.
Therefore, the choice of parallelism strategies not only affects local computation but also determines the global communication pattern.

Fig. \ref{fig:motivation}(b) shows a simplified Transformer layer, where an input tensor is first multiplied by a parameter matrix, then processed by a self-attention block, and finally multiplied by another parameter matrix. 
By adjusting the parallelism strategies assigned to the two MatMuls, we obtain two widely used intra-operator parallelism plans. 
On the left, both MatMuls adopt a row-wise parallelism strategy, allowing each device to independently compute on separate input data using a replica of the model parameters. 
During training, the gradients computed for parameters on each device must be synchronized to ensure consistent parameter updates across all replicas, thereby forming data parallelism.
On the right, the first MatMul adopts column-wise parallelism strategy, while the second MatMul uses split-K parallelism.
Each device produces a full-sized output tensor, but its element values are only partial results.
The partial results must be communicated immediately to assemble the full tensor before the next computation stage. 
This forms tensor parallelism plan, where communication is embedded within the forward and backward passes to maintain correctness.


Automatically generating such intra-operator parallelism plans is challenging, as it requires selecting a parallelism strategy for each operator, leading to a massive search space in large models.
This forces existing frameworks to build simplified symbolic cost models and adopt efficient search algorithms to find good solutions within a reasonable time.
For example, communication cost is typically estimated by the number of bytes derived from the shape and data type of the communicated tensors, while computation cost is measured by the number of floating-point operations required by each operator.



In practice, there is a substantial gap between the cost estimated by the symbolic model and the actual performance produced during execution.
Consider data and tensor parallelism for the transformer layer in Fig.\ref{fig:motivation}(b).
Given a \textit{hidden\_size} of 5120, \textit{seq\_len} of 1024, and \textit{batch\_size} of 16, Alpa~\cite{zheng2022alpa} estimates the cost of each parallelism plan by theoretically analyzing the communication data volume on the computation graph. 
It produces cost values of $1.51 \times 10^9$ for the tensor parallelism plan and $1.89 \times 10^9$ for the data parallelism.
However, the actual communication overhead diverges significantly from the theoretical communication volume-based cost. 
As shown in Fig.~\ref{fig:motivation}(c), tensor parallelism incurs extra overhead due to a compiler restriction that forces RNG operators to run on a single GPU. This requires an additional communication step to broadcast random data to other GPUs.
In contrast, data parallelism benefits from fusing communication for gradients of multiple parameters into a single, more efficient All-Reduce.
On 4 A100-PCIe GPUs, the communication time of data parallelism is only 60\% of that of tensor parallelism, despite its higher theoretical cost. This highlights the need to estimate cost based on the physical execution plan produced by the compilation and runtime system.

This gap arises because existing frameworks must efficiently search over a massive parallelism space, which forces them to rely on static analysis on the computation graph.
We observe that these frameworks overlook two types of important structural information, which are crucial for reducing search complexity. 
By exploiting these structural properties, we can extract a small set of representative parallel programs and profile them to estimate the costs of physical execution plans.
To explore this idea, we develop \sysname, a profile-based automatic intra-operator parallelism search system.

\vspace{-1mm}
\subsection{System Overview}

\begin{figure}
    \centering
    \includegraphics[width=0.48\textwidth]{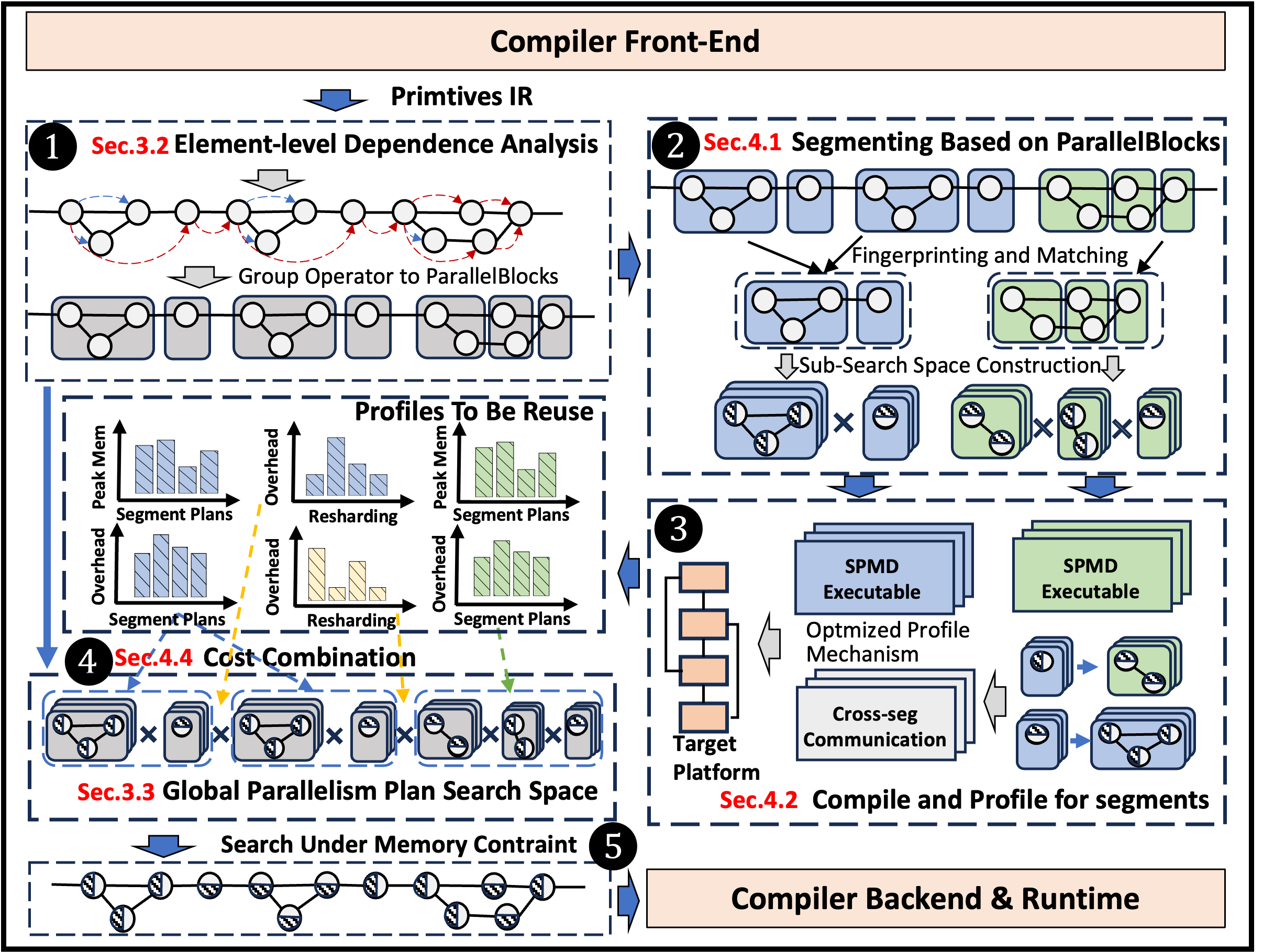}
    \vspace{-7mm}
    \caption{System Overview}
    \vspace{-5mm}
    \label{fig:overview}
\end{figure}

\sysname performs its analysis and optimization on the XLA compiler’s intermediate representation. 
It uses the compiler frontend to convert the model into a computation graph, and uses the backend and XLA runtime to generate and profile the representative programs. 

Fig.~\ref{fig:overview} shows the system overview of \sysname. 
\textbf{\textcircled{1}} After the compiler frontend translates the model into a static single assignment (SSA) form, it produces a linearized sequence of primitive operators.
\sysname analyzes element-level data dependencies between operators to identify subgraphs with parallelism-preserving properties. 
It then groups the corresponding operators into {\pb}s, which serve as basic units for constructing the parallelism search space, i.e., combining the parallelism strategies of multiple {\pb}s within the model or model segment to form a global or sub-parallelism search space.
\textbf{\textcircled{2}} \sysname treats the entire graph as a sequence of {\pb}s and applies template matching to extract a set of distinct subsequences (model segments). 
It extracts the element-level dependency graph of tensor contraction operators from the subsequences, which is used as a fingerprint for matching with other subsequences. 
This approach ensures that the matched subsequences share the same sub-parallelism space and parallel execution behavior, while significantly reducing the complexity of segment matching.
\textbf{\textcircled{3}}\sysname profiles each distinct model segment separately by building and running all plans in its sub-search space, collecting the computation and communication overhead, as well as memory consumption for each plan.
In addition, it also profiles the cost of tensor resharding across neighboring {\pb} pairs when there are mismatches between the output tensor partition of one segment and the input partition expected by the next.
\textbf{\textcircled{4}} It reuses the profiling results to estimate the execution cost of all matched segments, as well as the overhead of cross-segment tensor resharding. 
Subsequently, it aggregates these estimates to compute the overall execution cost and memory consumption of each global parallelism plan.
\textbf{\textcircled{5}} It uses a dynamic programming algorithm to find the optimal parallelism plan while satisfying the memory limit of the target platform.
The plan is then passed to the compiler backend to generate the final intra-operator parallel program.

%% file: content/Method/ParallelBlock.tex
\section{ParallelBlock Construction}\label{chap3_label}


\subsection{Parallelism-preserving Subgraphs}\label{sec_pb_def}

Computation graphs inherently contain many subgraphs, such as chains of element-wise operators, which maintain the parallelism of input tensors throughout their computation.  
In these subgraphs, regardless of the dimension along which an input tensor is partitioned, the computation can proceed independently on each parallel partition from input to output.
We refer to such subgraphs as parallelism-preserving subgraphs.

\begin{figure}
    \centering
    \includegraphics[width=0.48\textwidth]{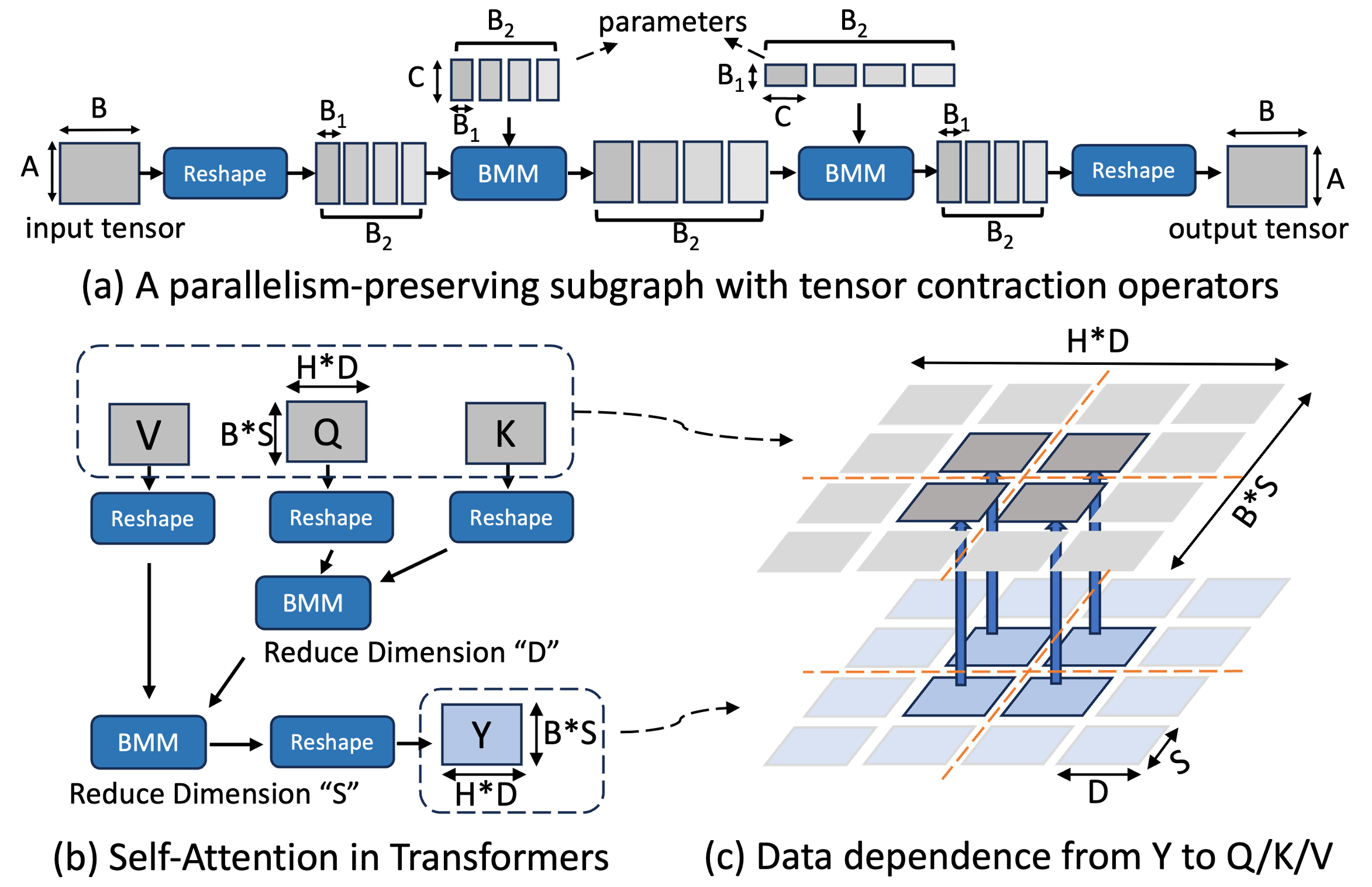}
    \vspace{-8mm}
    \caption{Parallelism-preserving examples. $B$, $S$, $H$, and $D$ in Self-Attention represent \textit{batchsize}, \textit{sequence length}, \textit{number of attention heads}, and the \textit{hidden size} of each head, respectively.  The computation graph is simplified for clarity.}
    \vspace{-4.5mm}
    \label{fig:parallelism-preserving_def}
\end{figure}

Besides simple chains of element-wise operations, a parallelism-preserving subgraph may also include data manipulation and tensor contraction operators.
For example, in Fig. \ref{fig:parallelism-preserving_def}(a), an input tensor with shape \([A, B]\) is processed by a sequence of reshape and batched matrix multiplication (BMM) operators, producing an output tensor with the same shape.
The input tensor is first reshaped to a tensor of shape \([A, B_1, B_2]\), by splitting the second dimension of the original tensor into two dimensions of sizes \(B_1, B_2\), respectively, where \(B_1 \times B_2 = B\). 
This tensor is then multiplied with a parameter of size \([B_2, C, B_1]\) using BMM operator, contracting the dimension with size \(B_1\) and introducing a new dimension with size \(C\), resulting in an intermediate tensor of size \([B_2, A, C]\).  
Next, a second BMM with a parameter of shape \([B_2, B_1, C]\) further contracts the dimension with size \(C\) and restores the tensor to shape \([A, B_1, B_2]\).  
Finally, a reshape operation merges two dimensions with sizes \(B_1\), \(B_2\), producing an output tensor with same shape as input tensor.

In this subgraph, the input tensor can be partitioned along the row dimension (size A), the column dimension (size B), or both.
Suppose the column dimension is mapped to an axis of the device mesh consisting of $P$ devices. 
We can observe that, if \(B_2\) is divisible by $P$, all tensor contractions in this subgraph can be performed independently within each parallel partition, without requiring inter-device communication.
Thus, the computation can proceed independently on each parallel partition from input to output.
This condition is commonly satisfied in practice, as intra-operator parallelism typically maps tensor dimensions to mesh axes with small power-of-two sizes (e.g., 2, 4, or 8), while tensor dimensions such as \(B_2\) are usually larger power-of-two (e.g., 16, 32, or 64).


A more practical example is the self-attention block in the Transformer (Fig.~\ref{fig:parallelism-preserving_def}(b)), which provides a more intuitive illustration of the parallelism-preserving property.
Fig.~\ref{fig:parallelism-preserving_def}(c) illustrates the data dependency from its output tensor to its input tensors. 
Each “block” of elements in the output tensor $Y$ depends only on the corresponding “blocks” of elements in the input tensors $Q$, $K$, and $V$. 
That is, if the subgraph's computation is partitioned at the "block" granularity, the computation for each partition can be performed independently.
Similar to the subgraph in Fig.~\ref{fig:parallelism-preserving_def}(a), it can also be regarded as a parallelism-preserving subgraph as long as $B$ and $H$ are divisible by the number of devices along the corresponding mesh axis.

This parallelism-preserving property enables the construction of coarse-grained parallel units to reduce the intra-operator parallelism search space. 
However, identifying such subgraphs in the computation graph is highly challenging, as the computation graph only encodes tensor-level data dependencies. 
This makes it difficult to determine whether a subgraph’s output-to-input dependency stays within a parallel partition.  
As a result, prior work~\cite{zheng2022alpa, liu2023colossal} has relied on simple heuristics, such as merging operators based on their types.
However, such methods are insufficient to accurately identify parallelism-preserving subgraphs, especially those involving complex data manipulation and tensor contraction operators.


We address this challenge by modeling element-level data dependencies in the computation graph using affine transformation expressions, which enable the precise definition and identification of parallelism-preserving subgraphs.
Consider an output tensor $Y$ with shape $[A_0, A_1, ..., A_{n-1}]$, where $A_i$ denotes the size (the number of elements) of the $i$-th dimension. 
Each element of $Y$ is indexed as $\mathbf{a} = (a_0, a_1, \dots, a_{n-1})$, where $a_i \in [0, A_i - 1]$ is an index along the $i$-th dimension.
Similarly, for an input tensor $X$ with shape $[B_0, B_1, ..., B_{m-1}]$, we can express the data dependency from an element in $Y$ to elements in $X$ using affine mappings: 
$Y(\mathbf{a}) \rightarrow { X(\mathbf{b}) \mid \mathbf{b} \in \mathcal{F}(\mathbf{a}) }$.
where $\mathcal{F}(\mathbf{a}) = (f_0(\mathbf{a}), f_1(\mathbf{a}), ..., f_{m-1}(\mathbf{a}))$ specifies the set of input indices that $Y(\mathbf{a})$ depends on.
Each $f_k(\mathbf{a})$ is an affine function of $\mathbf{a}$ that may return either a single integer (a scalar index) or a set of integers (a set of element indices) along the $k$-th dimension of $X$.

\begin{table}[t]
  \centering
  \caption{Affine mappings for different type of operators. 
  The dimension specification indicates how tensor indices change based on the semantics of the operator. 
  }
  \vspace{-2mm}
  \adjustbox{max width=\linewidth}{
  \begin{tabular}{ccc}
    \toprule
    Type of Op & Dimension Specifications & Affine Mappings \\
    \midrule
    Elem-wise & / & $f_i(\mathbf{a}) = a_i$ \\
    Reshape & Split dimension $k$ to $k, k+1$ & 
    $\begin{cases}
      f_i(\mathbf{a}) = a_i & i < k \\
      f_k(\mathbf{a}) = K_{a+1} a_k + a_{k+1} \\
      f_i(\mathbf{a}) = a_{i+1} & i > k
    \end{cases}$ \\
    Reshape & Merge dimensions $k, k+1$ to $k$ & 
    $\begin{cases}
      f_i(\mathbf{a}) = a_i & i < k \\
      f_k(\mathbf{a}) = \left\lfloor a_k / B_{k+1} \right\rfloor \\
      f_{k+1}(\mathbf{a}) = a_k \bmod B_{k+1} \\
      f_i(\mathbf{a}) = a_{i-1} & i > k+1
    \end{cases}$ \\
    Transpose & Permute dimensions $i \rightarrow \pi(i)$ & 
    $f_i(\mathbf{a}) = a_{\pi(i)}$ \\
    Broadcast & Add dimension $k$ & 
    $\begin{cases}
      f_i(\mathbf{a}) = a_i & i < k \\
      f_i(\mathbf{a}) = a_{i-1} & i \geq k
    \end{cases}$ \\
    Contraction & Reduce dimension $k$ & 
    $\begin{cases}
      f_i(\mathbf{a}) = a_i & i \ne k \\
      f_k(\mathbf{a}) = \{ c \mid 0 \leq c < B_k \} \\
    \end{cases}$ \\
    \bottomrule
  \end{tabular}
  }
  \label{tab:deps_for_op}
  \vspace{-3.5mm}
\end{table}

Taking the dependency from a single operator's output tensor to its input tensor as examples, as shown in Table~\ref{tab:deps_for_op}. 
For a simple element-wise operator, each output element depends on the input element with the same index, i.e., $m=n$, and $f_i(a_0, ..., a_{n-1}) = a_i$, for $0 \leq i<n$.
For a reshape operator that splits the $k$-th dimension of the input tensor into two consecutive dimensions, the $k$-th and the $k+1$-th, in the output tensor.
If the original input dimension of size $B_k$ is split into two output dimensions of sizes $A_k$ and $A_{k+1}$, where $A_k A_{k+1}=B_k$.
Then each output element indexed by $\mathbf{a} = (a_0, \dots, a_{n-1})$ depends on the input element indexed by $\mathcal{F}(\mathbf{a}) = (f_0(\mathbf{a}), \dots, f_{m-1}(\mathbf{a}))$,
where:
(1) $f_i(\mathbf{a}) = a_i$ for $i < k$,
(2) $f_k(\mathbf{a}) = A_{k+1} a_k + a_{k+1}$, and
(3) $f_i(\mathbf{a}) = a_{i+1}$ for $i > k$.
That is, the dependent element's index along the $k$-th dimension is computed by combining indices of two split dimensions, while the indices for other input dimensions are either copied directly (for $i < k$) or taken from the corresponding shifted positions in the output (for $i > k$) due to the insertion of a new dimension.

We further characterize the property that the computation of each parallel partition can proceed independently without requiring elements from other partitions.
A subgraph can be regarded as a parallelism-preserving subgraph if there exists a set of positive integers $d_0, d_1, ..., d_{n-1}$, such that, for $0 \le i < n$, the affine mappings between any input tensor $X$ and any output tensor $Y$ satisfy:
\begin{align}
    f_{i}(\mathbf{a}) = \lfloor \frac{a_i}{d_i} \rfloor d_i + c: 
     \quad 0 \leq c<d_i, \quad \frac{A_i}{d_i}  \bmod  P_i = 0\label{eq:stop_grouping_criteria}
\end{align}
\noindent
where $P_i$ represents the number of devices along the device mesh axis to which the $i$-th tensor dimension is mapped.
This indicates that the input and output tensors of a parallelism-preserving subgraph have the same tensor dimensions, and the  dependency in each dimension remains within the corresponding parallel partition.

For a path in the computation graph, \sysname incrementally builds affine mappings for each operator along the path and composes them.
It starts by assigning identity affine mappings to a given input tensor. 
Then, for each subsequent operator, it builds specific affine mappings based on operator's type and dimension specification, as shown in Table.\ref{tab:deps_for_op}.
These affine mappings are composed with the accumulated mapping from previous operators. 
Through this recursive composition, \sysname maintains an up-to-date mapping relation that describes how each element of the current tensor depends on the original input tensor.

\vspace{-2mm}
\subsection{ParallelBlocks}\label{sec_group_op}

Intuitively, we can group operators in such subgraphs and apply consistent parallel dimensions throughout. 
This allows us to eliminate the need to exhaustively combine strategies for individual operators by pruning away suboptimal combinations. 
Motivated by this, \sysname constructs coarse-grained structures, \pb{}s, which serve as basic parallel units for intra-operator parallelization.
The construction of \pb{}s proceeds in two phases:

\textbf{Phase 1: Operator grouping}.
Since tensor contraction operators are more likely to cause dimension reductions or expansions that disrupt the parallelism-preserving property, \sysname treats each of them as a starting point for dependency analysis to incrementally identify the downstream parallelism-preserving subgraph.
Each tensor contraction operator is then grouped with all downstream operators within this subgraph.

The grouping process is illustrated in Algorithm \ref{BuildPB}.
Tensor contraction operators are first sorted by their depth in the computation graph.
For each one, \textit{DFSAndGroup} explores the downstream operators in a depth-first manner and groups those that belong to the same parallelism-preserving subgraph.
At each step, it checks the data dependency relationship using the condition stated in Eq.~\ref{eq:stop_grouping_criteria}.



\textbf{Phase 2: Refinement}.
Assume a globally optimal parallelism plan assigns inconsistent parallel dimensions to two operators located before and after a parallelism-preserving subgraph, respectively.
In this case, a tensor resharding is required somewhere between these two operators. 
The optimal plan may prefer to choose the smallest intermediate tensor produced between the two operators for resharding to minimize communication overhead.
If such a tensor is part of the parallelism-preserving subgraph, treating this subgraph as a unit for parallelization may hinder internal resharding opportunities.
More specifically, there are two typical patterns that can lead to such internal resharding opportunities: (1) Multiple tensor contractions, where an input tensor dimension is first split into two dimensions, one of which is contracted and later projected back to the original shape. (2) Complex data manipulation operators (slicing or gather operations with certain patterns) that extract subsets or supersets of the input tensor and are later broadcast or reduced to match the expected output.

\setlength{\textfloatsep}{5pt}
\begin{algorithm}[t]
\small
\caption{DFS for Constructing ParallelBlock. }
\label{BuildPB}
\Input{ Computation Graph $G$}
\Output{ ParallelBlocks $PBS$: Lists of operators }
\SetKwFunction{MainFunc}{BuildParallelBlocks}
\SetKwFunction{SubFunc}{DFSAndGroup}
\SetKwFunction{SortTensorContractionOpSet}{SortTensorContractionOpSet}
\SetKwFunction{IsGrouped}{IsGrouped}
\SetKwFunction{SetGrouped}{SetGrouped}
\SetKwFunction{GetAllUsers}{GetAllUsers}
\SetKwFunction{AddOtherOperandToPB}{AddOtherOperandToPB}
\SetKwProg{Fn}{Function}{:}{\KwRet} 
\Fn{\MainFunc{$G$}}{
$PBS$ = \{\} \;
$S$ = \SortTensorContractionOpSet{$G$}\;
\For{each $s$ in $S$}{
    $PB$ = \{\} \;
    \If{\IsGrouped{$s$}}{
        continue\;
    }
    \Else{
        Add $s$ to $PB$\;
        \SubFunc{$s, PB$}\;
        \SetGrouped{$PB$}\;
    }
    Add $PB$ to $PBS$\;
}
\KwRet{$PBS$}\;
}
\Fn{\SubFunc{$op, PB$}}{
$users$ = {\GetAllUsers{$op$}}\;

\For{each $user$ in $users$}{
        \If{\IsGrouped{$user$}}{continue\;}
        \Else{
            \If{Check $user, PB$ with Eq.(\ref{eq:stop_grouping_criteria})}{
                Add $user$ to $PB$\;
                \KwRet{\SubFunc{$user, PB$}}\;
            }

        }
}
\KwRet{}\;
}
\end{algorithm}

Based on this, 
\sysname identifies
each parallelism-preserving subgraph that contains multiple tensor contractions or complex data manipulation operators, where an intermediate tensor is smaller than the subgraph’s output. \sysname then splits it at the corresponding resharding point, yielding two separate subgraphs.

\textbf{\pb}.
We define each subgraph constructed through the two-phase process as a \pb{}.
A \pb is a subgraph within which no internal resharding opportunity exists.
The parallel dimensions of operators in the \pb are determined by those of its first operator. 

\begin{theorem}
Enforcing consistent parallel dimensions throughout a \pb does not exclude the globally optimal parallelism plan from the reduced search space.
\end{theorem}

\textbf{Proof sketch.}
Consider a parallelism-preserving subgraph in the computation graph and a globally optimal parallelization plan.
There are two cases:

\textbf{Case 1:} The optimal plan assigns \emph{consistent} parallel dimensions to the operators located immediately before and after the subgraph. In this case, using parallel dimensions consistent with those of the preceding operator throughout the whole subgraph is necessary. Otherwise, it would introduce additional tensor resharding within the subgraph, which cannot be part of the optimal plan.

\textbf{Case 2:} The optimal plan assigns \emph{inconsistent} parallel dimensions to the operators located immediately before and after the subgraph.
In this case, tensor resharding is inevitable.
Since the parallelism-preserving subgraph in \pb is refined to exclude potential internal resharding opportunities, performing the resharding on the subgraph’s output tensor would not introduce additional communication overhead in comparison to resharding any intermediate tensors within the subgraph.

In both cases, enforcing consistent parallel dimensions throughout a \pb does not exclude the globally optimal parallelization plan.$\square$

\textbf{Construction complexity.}
Given a computation graph contains $N$ operators and $E$ edges. Let $T$ be the number of tensor contraction operators, and $K$ be the number of \pb{}s requiring intermediate tensor checks ($K < T$). Assume each such \pb{} contains on average $L$ edges.
The construction involves three steps:
(1) Sorting the $T$ contraction operators based on topological depth, which requires $O(N + E + T \log T)$ time.
(2) Constructing each \pb using DFS traversal, taking $O(N + E)$ as each operator and edge is visited at most once, to construct the affine transformation and compose it with previous accumulated dependency;
(3) For $K$ \pb{}s, checking edges (intermediate tensors) to find potential cheaper resharding points, with $O(K \cdot L)$ complexity.
Overall, the total time complexity is $O(N + E + T \log T + K \cdot L)$, which approaches $O(N + E)$ in practice since $T \ll N$ and $K \cdot L < E$.

\vspace{-1mm}
\subsection{Parallelism Plan Construction}\label{SearchSpace}                                     
To generate an intra-operator parallelism plan for computation graph, each operator should be assigned a parallelism strategy. 
\sysname first selects a parallelism strategy for the first operator of each \pb, and then automatically infers the parallelism strategies for all operators in the computation graph.

As shown in Fig.\ref{fig:parallel_plan}(a) to \ref{fig:parallel_plan}(b), the parallelism strategies of the operators within the two {\pb}s can be inferred by propagating the parallel dimensions of the first operator.
Operators in model parameter input branches are not grouped into {\pb}s because they lie outside the traversal path in \pb construction process. 
\sysname finds the operator within the \pb that receives this input branch and propagates its parallel dimensions back to the input branch. 
In Fig.\ref{fig:parallel_plan}(b), the parallel dimensions of the right side parameter are determined based on the parallel dimensions of operators in the second \pb.

In addition, some operators may generate tensors used by multiple {\pb}s, and the parallel dimension requirements of different {\pb}s for these operators may conflict. 
\sysname identifies the data dependency paths from these operators to all related {\pb}s and determines compatible parallelism strategies for them, as shown in Fig. \ref{fig:parallel_plan}(c).

Based on the parallelism plan construction process, \sysname builds the intra-operator parallelism search space by combining the parallelism strategies of all {\pb}s on computation graph, and then constructs a parallelism plan for each instance within the space.
The intra-operator parallelism search space is reduced to a combination of parallelism strategies for a few key operators instead of all operators, 
i.e., for a part of the model with $N$ {\pb}s and $D_i$ parallelism strategies for the i-th \pb, 
the size of the parallelism search space is $S = \prod_{i=1}^{N} D_i$.

%% file: content/Method/Profile.tex
\section{Profiling Model Segments}\label{chap4_label}

\subsection{Segment Generation}
\sysname identifies repeated computational patterns in the model by extracting a small number of distinct segments from the computation graph. 
Each type of segment shares the same parallelism search space and exhibits similar parallel execution behavior, allowing its physical execution cost to be profiled once and reused across all instances.
This extraction is achievable due to the graph's representation based on \pb. 
Instead of matching arbitrary operator-level subgraphs, \sysname uses \pb{}s as the matching granularity, turning segment identification into a sequence matching problem over \pb{}s.

However, directly comparing all operators within \pb{} sequences is impractical for two main reasons. First, each \pb{} typically contains a large number of operators, making exhaustive operator-level comparison computationally expensive. 
Second, strict matching at the operator level can lead to mismatches between semantically equivalent segments due to trivial variations. 
For example, one model layer may explicitly reshape a parameter tensor before use, while another functionally identical layer may consume the same tensor directly without reshaping. 
Such discrepancies do not affect the semantics or parallel execution behavior of the segment but can disrupt matching.

\begin{figure}
    \centering
    \includegraphics[width=0.45\textwidth]{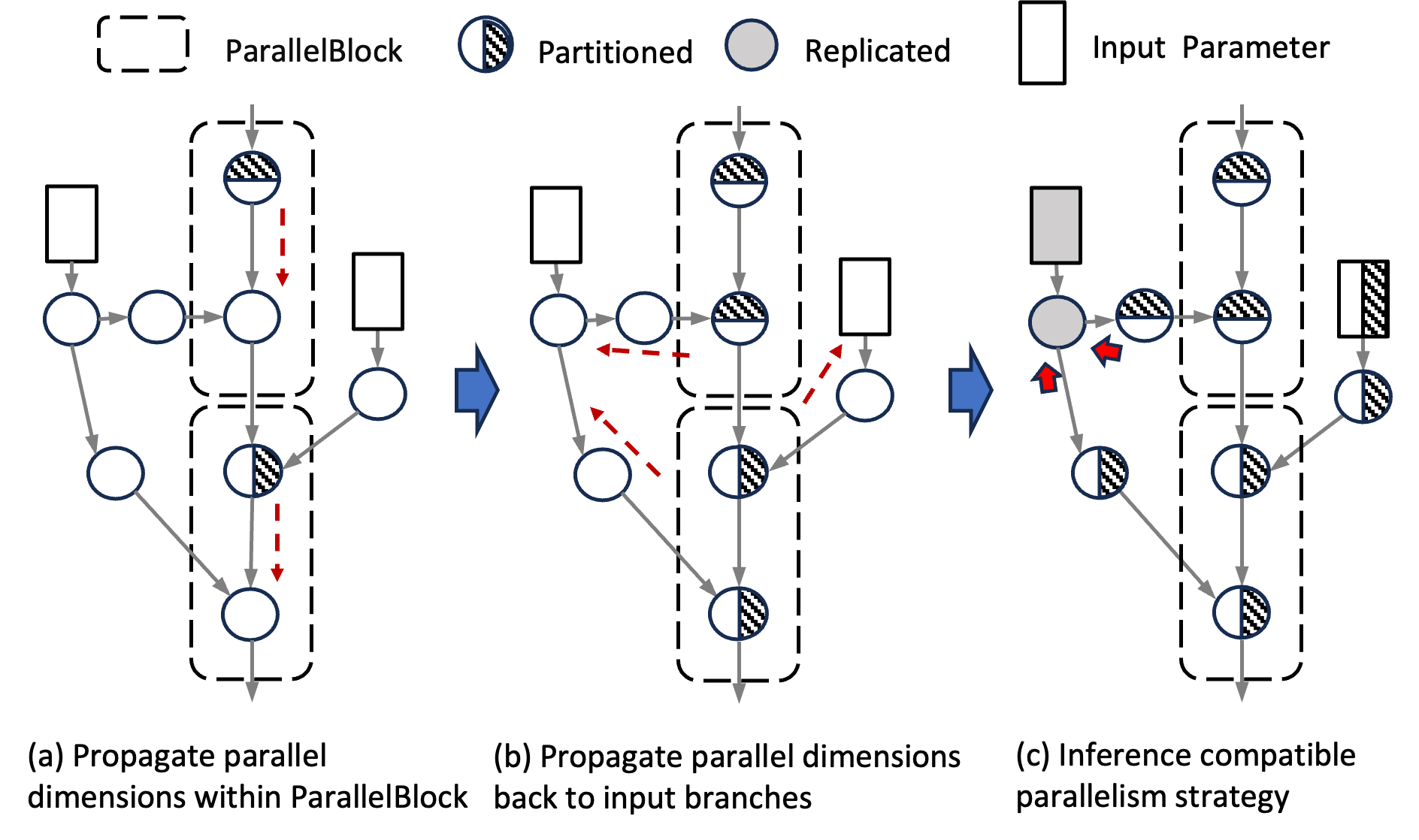}
    \vspace{-3mm}
    \caption{Inferring parallelism plan for a given parallelism strategy for each \pb.}
    \label{fig:parallel_plan}
\end{figure}

To address this, \sysname first searches for pairs of \pb{} subsequences that have the same key operators (i.e., the first operator in each \pb).
Since these operators dominate the choice of parallelism strategies of the subsequence, matching them ensures that the subsequences share the same parallelism search space.
To further determine whether these subsequences exhibit similar physical execution behaviors, \sysname extracts the data dependency graphs of tensor contraction operators from each subsequence as \textit{fingerprints} for comparison.
In this data dependency graph, each node represents a tensor contraction operator, and each edge is annotated with an element-level affine mapping that encodes how the output elements of one tensor contraction operator are consumed by another.
The construction of these affine mappings follows a similar process in Section~\ref{chap3_label}, i.e., composing the affine mappings of each operator along the dependency path.

\sysname then leverages this matching mechanism to perform template matching over the \pb{} sequence, aiming to identify a minimal set of representative subsequences that collectively cover the entire model. It builds a match table by comparing the fingerprints of all candidate subsequences and employs a dynamic programming algorithm to select the smallest set of segments with full coverage.

The rationale of the design of fingerprints is twofold.
\textbf{Ensuring similar execution behaviors.}
Tensor contraction operators typically dominate the computational cost, while other operator types incur negligible overhead.
Therefore, matching contraction patterns guarantees similar computation workloads.
In addition, aligning the element-level data dependencies among tensor contraction operators ensures that the same parallel strategy produces consistent intermediate tensor layouts and resharding behaviors, resulting in identical communication patterns.

\textbf{Improving matching efficiency.}
Fingerprint-based comparison not only avoids the high cost of exhaustive operator-level matching but also abstracts away trivial variations, such as explicit reshaping, that would disrupt matching.

\begin{figure}
    \centering
    \includegraphics[width=0.40\textwidth]{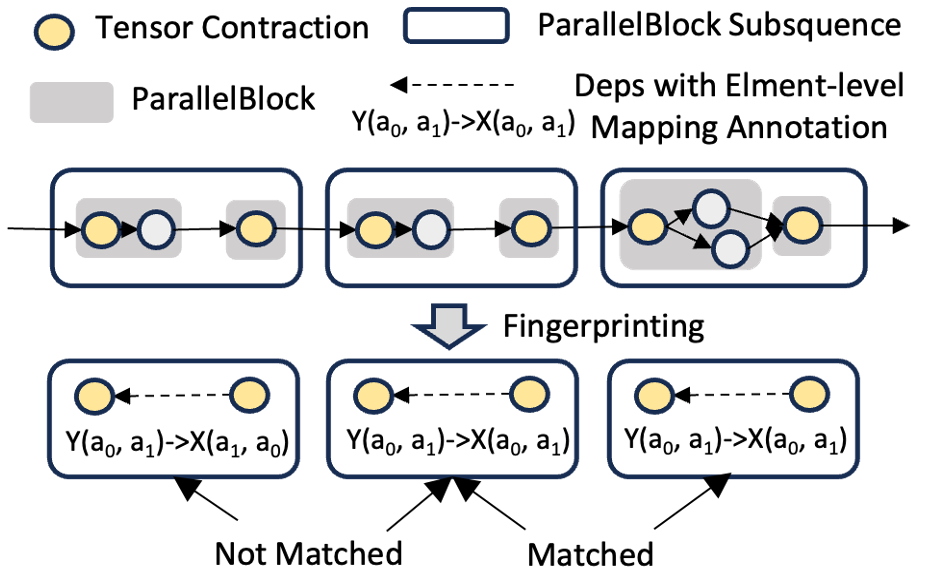}
    \vspace{-4mm}
    \caption{Identify distinct model segments based on \pb subsequences using element-level dependency graph of tensor contraction operators within \pb{}s.}
    \label{fig:model_segment}
    \vspace{-1mm}
\end{figure}

Fig.\ref{fig:model_segment} shows a simple \pb sequence containing three subsequences, each with two {\pb}s. 
Although the first and second subsequences contain the same tensor contraction operators, they differ in how intermediate results are consumed.
As a result, such differences in data dependencies may lead to divergent communication patterns under identical parallelism plans. For example, one subsequence may require additional resharding of intermediate results while the other does not.
In contrast, despite minor differences in non-tensor contraction operators, the last two subsequences have identical fingerprints, thus exhibit similar execution behavior under the same plans.

\vspace{-2mm}
\subsection{Profiling Mechanism}\label{profling_method}

\sysname first combines all {\pb}s' strategies to construct a sub-search space for each distinct segment.
It then profiles the communication and computation overhead for each parallelism plan in the sub-search space.
Profiling communication overhead is essential as it varies significantly across different plans.
Computation overhead is also profiled because different ways of partitioning operators can affect device utilization of computation kernels.

Directly aggregating the profiled results of all segments does not fully capture the cost of a global parallelism plan, because cross-segment tensor dependencies may introduce additional resharding overhead when the connected segments adopt incompatible strategies.
A straightforward solution is to enumerate all connected distinct segment pairs and profile the resharding overhead under every combination of their parallelism plans. 
However, this leads to a combinatorial explosion, as the number of combinations grows exponentially with the number of distinct segments and plans.

\sysname identifies cross-segment tensor dependencies in the computation graph and locates the corresponding source and destination \pb{}s involved in each dependency.
It then profiles the resharding overhead for these \pb{} pairs under all combinations of their parallel strategies.
This significantly reduces the profiling space while still capturing the cross-segment resharding cost needed for accurate modeling.
In practice, most cross-segment dependencies occur at segment boundaries, where the output tensor of one segment is directly consumed as the input tensor of the next.
Therefore, the profiling space for cross-segment tensor resharding is much lower than that for individual segments.


Overall, \sysname leverages the compiler backend to generate SPMD programs for all parallelism plans of each distinct segment, as well as for all cross-segment resharding cases. It then runs these programs multiple times to collect two types of profiles for all distinct segments:
(1) communication kernel time, computational kernel time, and peak memory consumption under different parallelism plans;
(2) communication kernel time for tensor resharding between connected cross-segment \pb pairs.

\vspace{-2mm}
\subsection{Profiling Overhead Analysis}

Assume that a model has $M$ {\pb}s and $N$ distinct segments, the $i$-th distinct segment contains $K_i$ {\pb}s, and the $j$-th \pb within this segment has $S_{i,j}$ parallelism strategies. 
We model the size of profiling space, i.e., number of parallel programs that need to be compiled and profiled as:

\vspace{-1mm}
\begin{align}
    \sum_{i=1}^{N}{(\prod_{j=1}^{K_i} S_{i,j} + \sum_{j=1}^{N} \sum_{(l, k) \in D_{i,j}}{S_{i,l}*S_{j, k}})} 
\end{align}
\noindent
where $D_{i, j}$ is a list containing pairs of index $(l, k)$ of {\pb}s that have cross-segment tensor dependencies within segment $i, j$.
The left side of the summation represents the number of parallelism strategies combinations for all {\pb}s within a segment, while the right side is the number of parallelism strategy combinations for cross-segment dependent \pb pairs.

In the worst case, where the entire model acts as a single distinct segment, the number of programs that need to be compiled and profiled is $\prod_{j=1}^{M}{S_{j}}$, which is the simple combination of the parallelism strategies of $M$ {\pb}s, as explained in Section~\ref{SearchSpace}.
In the best case, where the model consists of a single distinct segment repeated, and the cross-segment tensor dependencies only occur between the first and last \pb within it, the profiling space is $\prod_{i=j}^{K} S_{j} + S_{1}*S_{K}$, i.e., the sum of all parallelism strategies for the K {\pb}s within a single segment, plus the resharding from the last \pb to the first \pb.

Thanks to the extensive use of repeated model structures and wide range of parallelism-preserving subgraphs in existing models,  
the number of distinct segments and the {\pb}s per segment remains within a reasonable range.
For example, after grouping two BMM operators in self-attention into a single \pb, the repeated transformer layer in GPT model has only 4 {\pb}s, corresponding to four matrix multiplication operators.
As a result, we can evaluate the costs of all global parallelism plans by profiling just a few hundred short parallel programs.

Running programs multiple times incurs a significant time overhead, depending on the training workload and the execution efficiency of the parallelism plans.
In addition, the compilation time for parallel programs can also be substantial, especially when the computation graph is long but training workload is light-weight.
To further mitigate the overhead, \sysname introduces two optimizations:
First, it enforces a dynamic profiling time budget, which is continuously updated based on the fastest observed parallelism plans, aggressively trimming the profiling of inefficient or stalled executions.
Second, it parallelizes the compilation of different programs and overlaps the compilation phase with the profiling phase.

\vspace{-1mm}
\subsection{Plan Cost Estimation and Optimization}
We represent a global parallelism plan as a tuple $\mathbf{i} = (i_1, i_2, \dots, i_N)$, where $i_n$ denotes selecting the $i_n$-th parallelism plan for the $n$-th segment in a model with $N$ segments.
We estimate its execution time $T(\mathbf{i})$ and memory consumption $M(\mathbf{i})$ using profiles of distinct segments.
For the $n$-th segment that selects the $i_n$-th parallelism plan, we denote $p_n(i_n)$ as the computation time, $c_n(i_n)$ as the communication time, and $m_n(i_n)$ as the peak memory usage.
For each adjacent pair of segments $(n, n+1)$, the resharding cost when using the $i_n$-th and $i_{n+1}$-th parallelism plan is denoted as $r_n(i_n, i_{n+1})$.
The total execution cost and memory usage are defined as:

\begin{equation}
T(\mathbf{i}) = \sum_{n=1}^{N} (p_n(i_n) + c_n(i_n)) + \sum_{n=1}^{N-1} r_n(i_n, i_{n+1})
\end{equation}

\begin{equation}
M(\mathbf{i}) = \sum_{n=1}^{N} m_n(i_n)
\end{equation}

The first term in \( T(\mathbf{i}) \) aggregates segment computation and communication costs, while the second term sums the cross-segment resharding overheads.
Each part of the cost calculation is extracted from the profile of distinct segments, ultimately combining to represent the overall cost of the global parallelism plan $\mathbf{i}$.



\sysname searches for the plan with the minimal execution cost under the constraint of the memory usage limit, which is formulated as a dynamic programming problem.
It first builds a cost table and a memory table for each segment’s possible plans, based on the profiles of the distinct segments.
Then, starting from the first segment in the computation graph, the algorithm recursively explores all feasible combinations of segment plans, accounting for segment cost, cross-segment resharding cost, and accumulated memory consumption.
To reduce the complexity of the dynamic programming, we quantize the memory usage of each parallelism plan.

This segment-based cost combination method not only reuses segment costs to reduce profiling overhead, but also allows flexible parallelism plan selection.
Segments (even those of the same type) may adopt different plans: some high-throughput but memory-intensive, others low-throughput but memory-efficient, maximizing overall throughput while staying near the memory limit.


%% file: content/Evaluation/Setup.tex
\vspace{-1mm}
\section{Evaluation}

\begin{figure*}[ht]
    \centering
    \includegraphics[width=\textwidth]{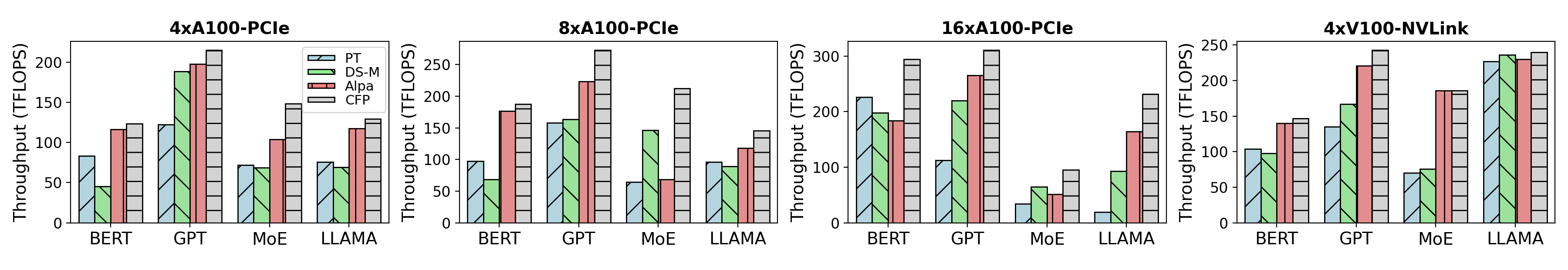}
    \vspace{-10mm}
    \caption{The average training throughput achieved by four parallel frameworks on two platforms.}
    \label{fig:e2e_perf_1201}
    \vspace{-5mm}
\end{figure*}

\sysname is an optimization system based on the Tensorflow-XLA framework.
We implemented optimization passes in the XLA compiler to build {\pb}s and extract distinct model segments. 
We also used the XLA compiler backend and runtime for generating and profiling SPMD programs. 

We compared \sysname with three parallel training frameworks: PyTorch (PT), DeepSpeed-Megatron (DS-M), and Tensorflow-Alpa (Alpa).
PyTorch employs data parallelism plans, which can be considered as intra-operator parallelism plans that split the batch dimension for all operators.
DeepSpeed-Megatron offers a manually designed parallel framework that incorporates both data and tensor parallelism. 
Alpa is one of the state-of-the-art automatic parallel frameworks that comprehensively considers inter-operator and intra-operator parallelism.
Its intra-operator parallelism search space contains parallelism plans equivalent to both data parallelism and tensor parallelism.
As Alpa's optimal solution for multi-level parallelism relies on the optimal intra-operator parallelism plan for each model segment (pipeline stage), we extracted its intra-operator parallel framework to compare with \sysname.
We believe that comparing with Alpa is appropriate, as its intra-operator search process has been widely adopted by newer parallel frameworks, such as nnScaler~\cite{nnscaler}, while also supporting more recent models than other automatic intra-operator parallel frameworks like FlexFlow~\cite{lu2017flexflow}.

\vspace{-2mm}
\subsection{Evaluation Setup}
We used four models in the evaluation: BERT~\cite{devlin2019bert}, GPT~\cite{gpt2020}, GShard MoE~\cite{GShardMoE} and LLAMA-2~\cite{touvron2023llama}. 
We set the micro batch size from 2 to 32 per GPU and report the average floating-point operations per second (FLOPS) during training.
\sysname can handle other models besides the evaluated ones because the construction of \pb{}s and distinct model segments is based on compiler IR.

During profiling of each distinct segment’s parallelism plan, we first ran the parallel program 5 times for warm-up, followed by 10 runs to collect the desired profile items. 
For the evaluation of the selected parallelism plan, we ran the test program 100 times and reported the average FLOPS.
We evaluated \sysname on two GPU nodes, each with 8 NVIDIA A100 40GB GPUs connected via PCIe. 
We also evaluated it on a node with 4 NVIDIA V100 16GB GPUs connected via NVLink to validate its optimization in the target platform with higher communication bandwidth.
On the A100-PCIe platform, all models are trained with TF32 precision, while on the V100-NVLink platform, we use FP16 precision due to memory constraints.

%% file: content/Evaluation/E2EPerf.tex
\vspace{-2mm}
\subsection{Training Performance}

Although \sysname significantly reduces the intra-operator parallelism search space, the reduced space still includes the data parallelism plan, the tensor parallelism plan, and the communication volume-optimal plans searched by Alpa. 
However, \sysname may not select these solutions as they might not achieve the best profile-based cost.
Fig. \ref{fig:e2e_perf_1201} shows the average throughput achieved by four parallel frameworks on two platforms. 
PT and DS-M utilize fixed parallel templates, which makes it challenging to achieve optimal throughput across all training settings. 
This limitation is even more significant for models that require more flexible parallelism plans, such as LLAMA and MoE models, leading to larger performance gaps between fixed parallel templates and optimized plans.

\sysname achieved an average performance speedup of 1.17x over Alpa with 4 GPUs and 1.63x with 8 GPUs.  
We observed that Alpa's communication volume-based cost model lacked awareness of downstream compilation and the efficiency of communication primitives, leading to inaccurate predictions of communication workload and efficiency in many experimental settings.  
For GPT and LLAMA, Alpa searched a tensor parallelism-like plan to minimize communication volume in transformer layers for small batch sizes.  
This introduced unexpected communication workloads during downstream compilation, offsetting the benefits of minimizing theoretical communication volume.  
In contrast, \sysname partitioned batch dimensions for each operator in these settings, achieving higher communication efficiency after All-Reduce kernel fusion and up to 1.51x and 1.31x speedup over Alpa on the two models, respectively.

For GShard MoE, \sysname achieved an average performance speedup of 2.13x over Alpa. 
Although Alpa's parallelism plan has the optimal communication volume, the compiled parallel program relies on multiple inefficient ncclSendRecv kernels on PCIe platforms. 
In contrast, \sysname used a hybrid parallelism plan: it splits batch dimensions for most operations but splits different dimensions for operators in the expert network, depending on the batch size. 
These parallelism plans rely on more efficient collective communication primitives, such as All-Gather and Reduce-Scatter.

\textbf{Multiple A100-PCIe Node.} 
We evaluated 4 frameworks on a 16-GPU cluster arranged as 1D and 2D device meshes (1×16, 2×8, 4×4), enabling hierarchical parallelism: one parallel dimension across nodes, another within nodes.
We used Alpa to explore the full space, while \sysname restricts the data batch dimension to the outermost mesh axis to keep the search space consistent with the 1D case.

A more complex communication topology makes it harder for Alpa's cost model to evaluate the performance of different parallelism plans accurately. 
It struggled to evaluate communication spanning multiple device levels, such as collective communication involving both inter-node and intra-node. 
As a result, even though its search space includes both 1D and 2D intra-operator parallelism plans, it overlooks 1D plans in many cases, even when those could provide better communication efficiency.

Moreover, multi-level communication provides the compiler with more optimization opportunities, making the actual communication workload deviate even more from the theoretical cost. 
For example, \sysname employs tensor parallelism for the expert network part in MoE. 
This requires All-Gather and All-Reduce for data aggregation before and after the expert network.
In practice, this plan results in less communication time because the compiler's downstream optimization automatically rewrite multiple communication kernels into more efficient ones with reduced communication volume.
However, Alpa overlooked this plan because its estimation of the resharding communication cost was 8x higher than the actual communication workload.
\sysname avoids these issues by profiling real-world communication costs, achieving a maximum speedup of 2.01x on BERT, 1.43x on GPT, 3.06x on MoE, and 1.67x on LLAMA.

\begin{figure}
    \centering
    \includegraphics[width=0.48\textwidth]{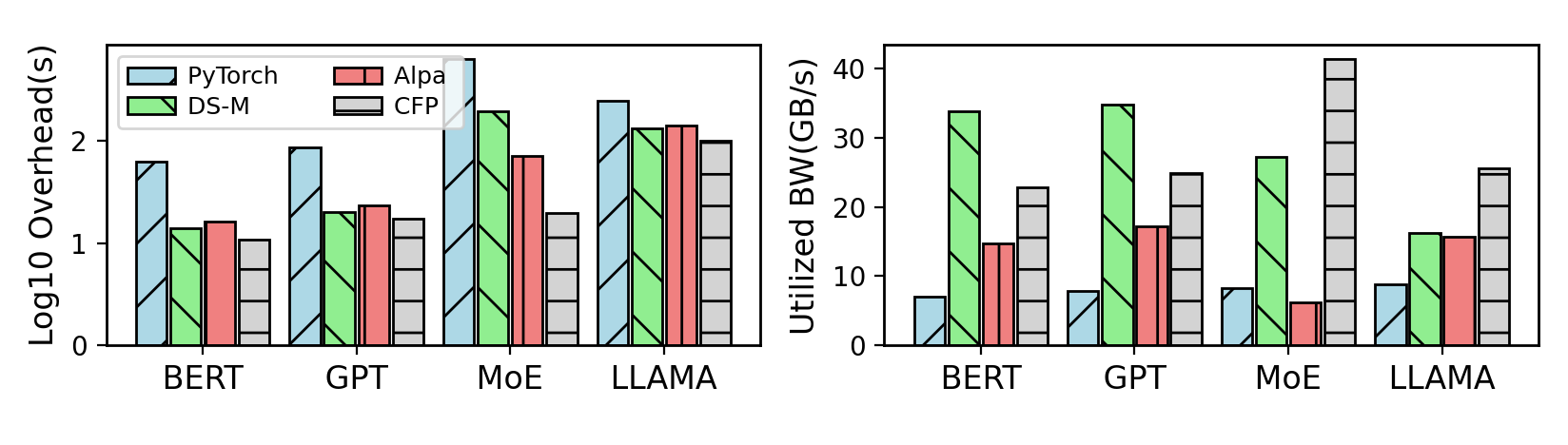}
    \vspace{-10mm}
    \caption{Overhead and achieved bandwidths of communication kernels for four models with batchsizes of 8, 8, 32, and 80, respectively. A logarithmic scale was used for communication overhead. }
    \vspace{-1mm}
    \label{fig:bandwidth}
\end{figure}

\textbf{Single V100-NVLink Node.} 
With the higher inter-GPU communication bandwidth provided by the NVLink platform, the proportion of communication time in the overall training time is reduced.
Since the main performance differences between parallelism plans arise from differences in communication patterns, the reduced communication overhead narrows the overall performance gap between different plans.
As shown in Fig.\ref{fig:e2e_perf_1201}, compared to PT, DS-M, and Alpa, \sysname achieved average speedups of 1.73x, 1.61x, and 1.05x.
For MoE, both \sysname and Alpa found plans that reduced communication time to about 1/20 of the total training time, leading to no significant performance differences.
For other models, Alpa applies the same parallelism plans as those selected for the PCIe platform.
In contrast, \sysname identified optimal parallelism plans based on the profiles from the NVLink platform.

%% file: content/Evaluation/Analysis.tex
\vspace{-4mm}
\subsection{Communication Overhead Analysis}

We observed that most parallelism plans evenly distribute heavy operators across multiple GPUs, making communication overhead the main factor causing significant differences in final throughput, especially on the A100-PCIe platforms.
Fig.\ref{fig:bandwidth} shows the communication overhead and utilized communication bandwidth by the four frameworks on 4xA100-PCIe GPUs. 
\sysname balanced communication volume and efficiency, achieving the smallest communication overhead across the four models.

\begin{figure}
    \centering
    \includegraphics[width=0.47\textwidth]{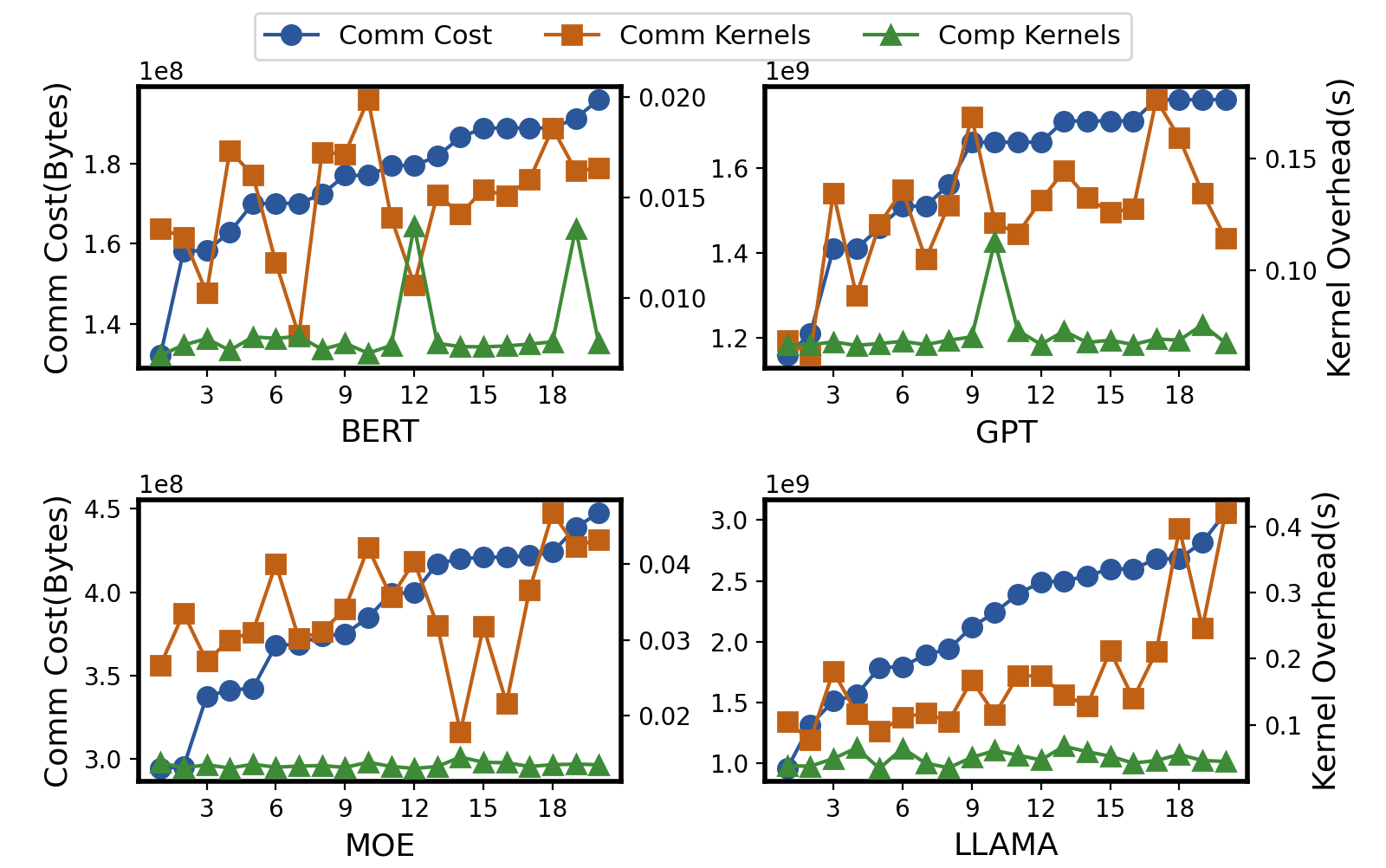}
    \vspace{-4mm}
    \caption{Time of computation and communication kernels for plans in \sysname's intra-operator parallelism search space. Sorted in ascending order of Alpa's communication volume-based cost, with only the top 20 selected for clarity.}
    \vspace{-1mm}
    \label{fig:volume_and_overhead}
\end{figure}

We further investigated the discrepancy between the theoretical cost based on communication volume and the measured performance. 
As shown in Fig.\ref{fig:volume_and_overhead}, 
while communication kernel overhead indeed increases with theoretical cost,
parallelism plans with similar theoretical cost can have significantly different communication overheads. 
For instance, in BERT, the 6th to 8th plans had similar theoretical costs, but their actual overheads differed by a factor of two. 
Moreover, plans with optimal communication overhead can have substantially higher theoretical costs, e.g., the parallelism plan (the 14th) with the highest efficiency in MoE had a theoretical cost 1.45 times that of the smallest cost.
We believe that two main reasons contribute to such a significant discrepancy:
(1) The unpredictable effects of code lowering and optimization: The compiler may generate unexpected communication kernels for certain communication patterns, or it may optimize the communication bandwidth through compiler optimization techniques. 
(2) The complex relationship between communication efficiency, target platform, communication algorithms, and tensor shapes: even with similar communication workloads, these factors can significantly affect the communication efficiency, leading to highly variable communication overhead.



%% file: content/Evaluation/MemConstraint.tex
\vspace{-1.8mm}
\subsection{Performance Under Memory Constraints}

\begin{figure}
    \centering
    \includegraphics[width=0.49\textwidth]
    {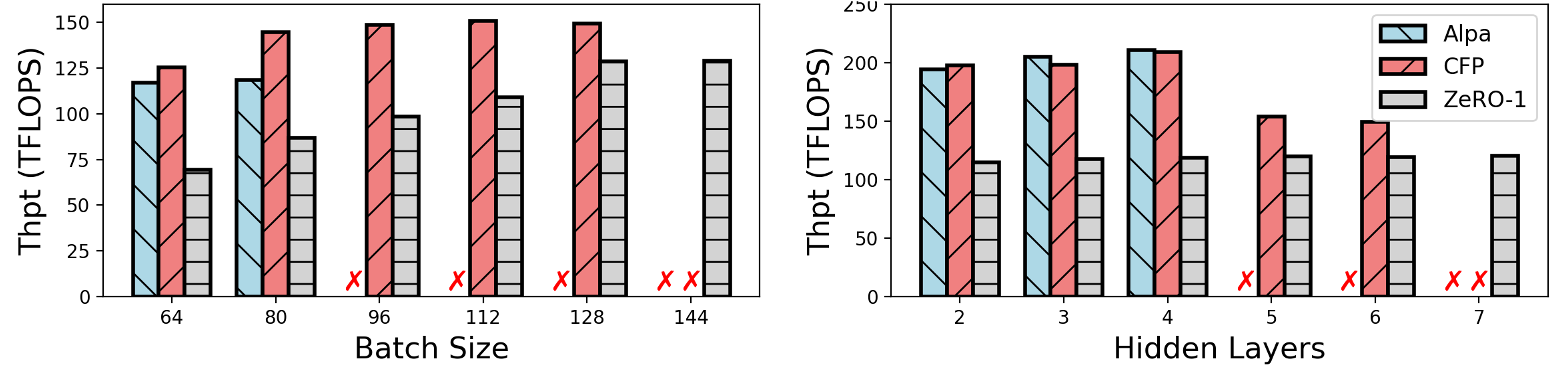}
    \vspace{-7.5mm}
    \caption{Training performance of LLAMA-7B with different hidden layer numbers and batch sizes. Left: Fixed hidden layer number to 6, increasing batch size. Right: Fixed batch size to 128, increasing the hidden layer number.}
    \label{fig:perf_mem_constraint}
    \vspace{-1mm}
\end{figure}

We evaluated the parallel training throughput of \sysname under memory constraints on four A100-40GB GPUs.
Fig.~\ref{fig:perf_mem_constraint} shows the throughput of different parallel frameworks for training the LLAMA2 model with varying numbers of hidden layers and batch sizes.
Alpa chose parallelism plans without integrating memory constraints into the search process, which quickly led to out-of-memory problems as batch sizes and the number of hidden layers increased. 
We also compared \sysname with ZeRO stage-1\cite{zero20sc}, a state-of-the-art memory optimization framework. 
It distributes all optimizer states across each GPU, reducing memory usage but suffering from high communication costs, which results in lower throughput.

\sysname balanced throughput and memory usage by selecting different parallelism plans for different segments, enabling the training of deeper models than Alpa while achieving higher throughput than ZeRO stage-1.
Note that \sysname inevitably prunes many memory-efficient plans, such as FSDP, which results in its memory optimization having a lower upper limit compared to ZeRO.

%% file: content/Evaluation/CompileOverhead.tex
\vspace{-1.8mm}
\subsection{Profiling Space and Search Overhead}
For BERT, GPT, and LLAMA, \sysname extracts two segments besides embedding/output layers: one for the first hidden layer and one for subsequent layers.
Although structurally identical, these segments have different fingerprints due to element-level dependency variations introduced during code lowering.
Each segment yields 4 {\pb}s. The first tensor contraction (matrix multiplication) in each has 3 candidate parallel dimensions, resulting in $3^4=81$ parallelism plans per segment. 
In addition, profiling is required for tensor resharding across segment boundaries and repeated segments, covering $3\times3=9$ groups of communication kernels. 
In total, $2 \times 81 + 2 \times 9 = 180$ programs are compiled and profiled.

For MoE models, \sysname treats alternating MoE and Transformer blocks as separate segments, each with 4 {\pb}s.
One \pb in MoE blocks involves batched matrix multiplication with an additional expert-related batch dimension, increasing its candidate dimensions and slightly expanding the profiling space.


\sysname's search overhead can be divided into four parts. 
\textit{AnalysisPasses}, refers to constructing {\pb}s and extracting distinct segments for model. 
\textit{ExecCompiling} represents the overhead of compiling and generates all executable programs that need to be profiled. 
\textit{MetricsProfiling}, refers to the overhead involved in running these programs multiple times to collect profiles.
\textit{ComposeSearch}, represents the cost of combining the profiling results of distinct segments to search for the globally optimal parallelism plan.

\begin{figure}
    \centering
    \includegraphics[width=0.48\textwidth]{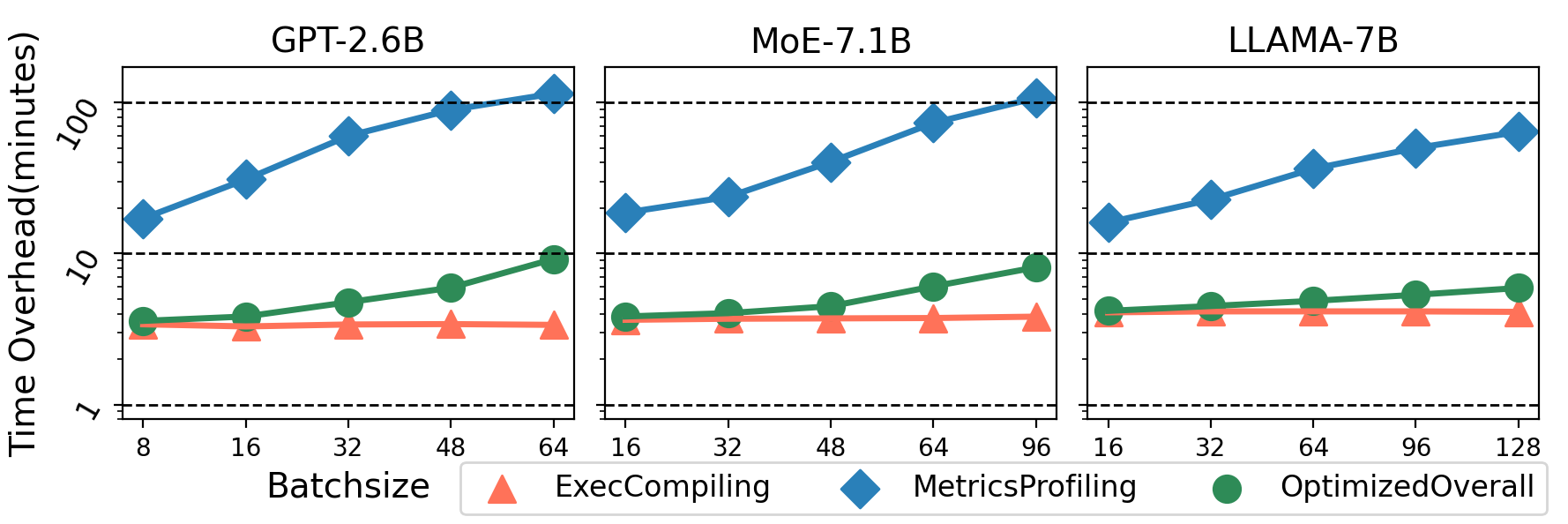}
    \vspace{-7.5mm}
    \caption{Compiling and profiling time for distinct segments in three models on a platform with 4 A100-PCIe GPUs.}
    \vspace{-5.5mm}
    \label{fig:search_overhead}
\end{figure}

The magnitude of \textit{ExecCompiling} depends on the number of distinct segments and the parallelism space of each segment. 
\textit{MetricsProfiling} is influenced by the training workload of each plan.
Fig. \ref{fig:search_overhead} shows these two overhead details for GPT-2.6B, MoE-7.1B, and LLAMA-7B. 
As the batch size increases, the \textit{ExecCompiling} remains relatively stable, while \textit{MetricsProfiling} increases. 
By parallelizing the compilation, overlapping it with the profiling process, and using a dynamic profiling time budget, \sysname significantly reduced the search overhead by trimming inefficient or stalled parallelism plans, as shown by \textit{OptimizedOverall} in Fig. \ref{fig:search_overhead}.

\begin{figure}
    \centering
    \includegraphics[width=0.48\textwidth]{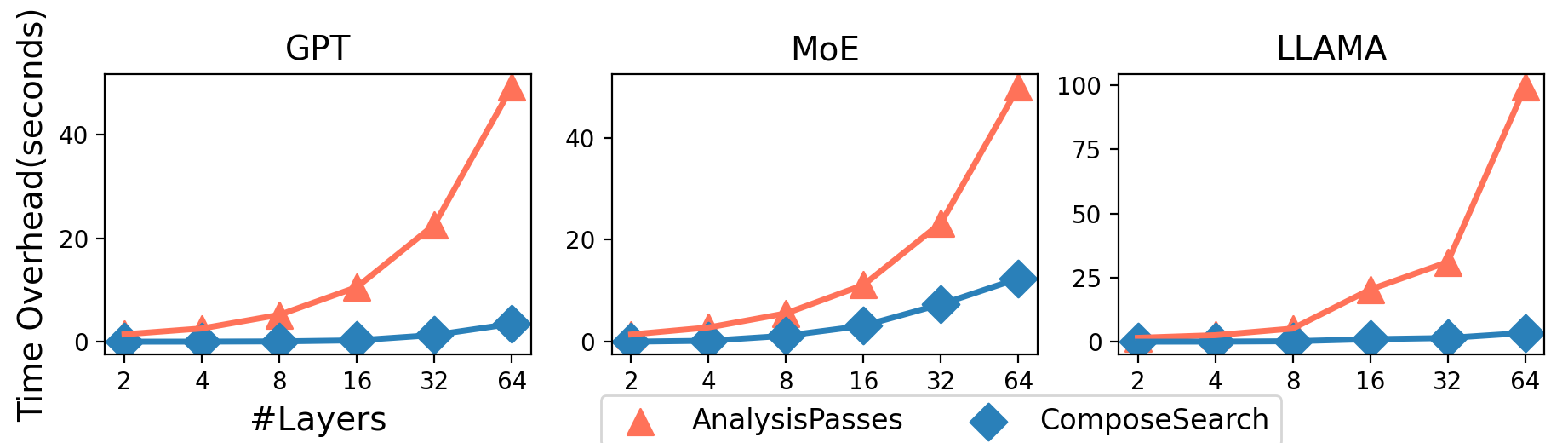}
    \vspace{-7mm}
    \caption{Analyzing and searching time for different numbers of layers of three models with a 24-core processor.}
    \vspace{-1mm}
    \label{fig:analysis_overhead}
\end{figure}

The overhead of \textit{AnalysisPasses} and \textit{ComposeSearch} is not affected by the training workload but increases as the model depth grows. We evaluated the overhead of them with different numbers of hidden layers on three models, as shown in Fig.\ref{fig:analysis_overhead}. In most intra-operator parallelism exploration scenarios (e.g., searching for parallelism plans inside a pipeline stage), their overhead is much smaller than \textit{ExecCompiling} and \textit{MetricsProfiling}.

\vspace{-1.8mm}
\subsection{Scalability Analysis}\label{scalability}

For larger systems with more GPUs, there are two typical cases where \sysname can be used with no increase in analysis and profiling complexity.
(1) Combine \sysname with data parallelism to exploit multi-dimensional intra-operator parallelism. Suppose the model is trained using data parallelism across multiple worker groups. \sysname can be applied in each group using its original workflow, and its profiles can be reused across all groups. 
(2) Combine \sysname with pipeline parallelism. \sysname can explore intra-operator parallelism within each potential pipeline stage, where the profile results of model segments (smaller than a stage) can also be reused for stage profiling to determine the best pipeline stage partitioning. 

For larger models, \sysname's profiling space will not increase unless there are new distinct segments. 
More layers may raise \sysname’s analysis and search overhead, and increasing configuration settings like hidden size may raise the profiling overhead for distinct segments. However, these overheads remain negligible compared to the training time.
While the complexity of the final search process increases nonlinearly with the number of layers, it is possible that future models with extremely deep architectures make dynamic programming infeasible. In such cases, techniques from database query optimization (e.g., genetic algorithms) can be employed while still leveraging physical execution costs to guide the search.


%% file: content/Evaluation/CaseStudy.tex
\begin{figure}
    \centering
    \includegraphics[width=0.47\textwidth]{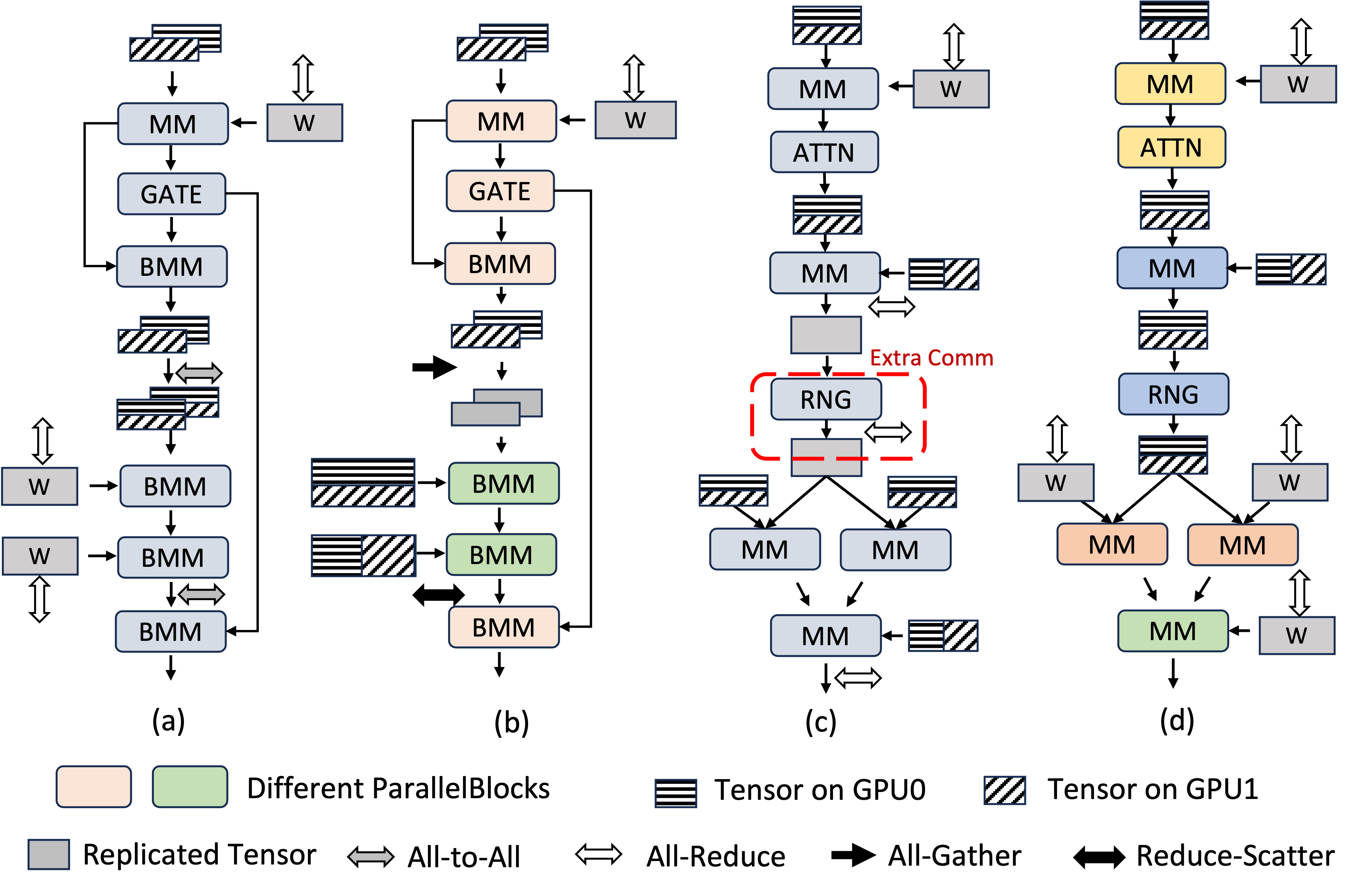}
    \vspace{-4.5mm}
    \caption{Parallelism plans searched by Alpa and \sysname for MoE-7.1B batchsize 16 and LLAMA-7B batchsize 80. GATE, ATTN, and RNG stand for gated network, attention block, and random number generator operator, respectively. }
    \vspace{-1mm}
    \label{fig:case_study_1205}
\end{figure}

\vspace{-3.5mm}
\subsection{Case Study}


\textbf{GShard MoE on A100-PCIe.}
Fig.12(a)(b) shows the parallelism plans searched by Alpa and \sysname for a single MoE layer. Alpa partitions the data dimension of each operator and adopts ``expert parallelism'' for the expert network.
This requires All-to-All communication to reshard intermediate results before and after the expert network and All-Reduce to aggregate gradients.
\sysname selects different plans for different batch size: when the batch size is smaller than 96, it uses a tensor parallelism-like plan to partition the parameters of the expert network. 
This requires All-Gather to reshard tensors before the expert network and All-Reduce to aggregate results, where the All-Reduce operation is rewritten by the compiler into a Reduce-Scatter.
For larger batch sizes, \sysname partitions the batch data dimension for all operators.
Alpa overlooked this plan because it  overestimated the communication cost of the tensor resharding after expert network by 8 times compared to the actual communication workload in 16 GPUs training. 
We observed that the parallelism plans searched by Alpa incurred twice the communication overhead of \sysname for batch sizes smaller than 96 and over three times the overhead for larger batch sizes.

\textbf{LLAMA on V100-NVLink.} 
Fig.12(c)(d) shows the parallelism plans searched by Alpa and \sysname for LLAMA-7B with a batch size of 80.
Alpa partitions the batch dimension of input data in the self-attention part and then partitions parameters for the remaining operators to minimize communication volume. However, this causes each GPU to call the random number generator operator independently, leading the compiler backend to insert extra All-Reduce for random data synchronization. 
In contrast, \sysname partitions all operators along the batch dimension, leverages the compiler to merge all gradient synchronizations into a single communication kernel for better efficiency.
Moreover, due to the higher interconnect bandwidth between GPUs on the NVLink platform, computation overhead has a more critical impact on overall training time. 
We observed that frequent communication in Alpa’s plan introduced more data movement operators (concat and split), making its computation overhead about 10\% higher than \sysname.

%% file: content/RelatedWorks.tex
\vspace{-3mm}
\section{Related Works}

\textbf{Modeling the Intra-operator Parallelism Search Space.}
Many works~\cite{xu2021gspmd, dtensor, yuan2022oneflow} propose concise yet expressive tensor partitioning methods that support diverse parallel paradigms. More recently, strategies that partition the temporal dimension have been used~\cite{summa_parallel, primpar_eurosys24, distal_pldi22, distribute_mapping_tensor_compute_sc23} to further expand the intra-operator parallelism space.

\textbf{Automatic Search for Intra-operator Parallelism.} 
To efficiently search for intra-operator parallelism plans, most works~\cite{tofu2019, cai2021tensoropt, pase, accpar, hap_heter_gpu_spmd_eurosys24, primpar_eurosys24}, have employed various symbolic cost models and dynamic programming-based search algorithms.
FlexFlow~\cite{lu2017flexflow} and Automap~\cite{schaarschmidt2021automap} used a Monte Carlo-based search algorithm and employed an ML-based approach to evaluate plans.
Colossal-Auto~\cite{liu2023colossal} and Alpa~\cite{zheng2022alpa} use an ILP solver to find the plan with the lowest communication volume.
Some works~\cite{aceso_eurosys24, uniap} also adopt profile-based approaches, profiling operators and communication primitives separately and aggregating the results into cost models.
However, this operator-level profiling approach still struggles to capture execution efficiency accurately.
Moreover, they mainly focus on stage-level pipeline partitioning and rely on predefined intra-operator parallel templates, rather than fully exploring the space.


\textbf{Combine Intra- and Inter-operator Parallelism.}
Many frameworks adopt fixed parallelism plans per pipeline stage, without exploring the intra-operator search space~\cite{narayanan2019pipedream, NEURIPS2021_piper, huang2019gpipe}.
Aceso~\cite{aceso_eurosys24} combines the exploration of intra-operator and pipeline parallelism by using reconfiguration mechanisms to alleviate performance bottlenecks.
Galvatron~\cite{galvatron} proposes a decision-tree abstraction to decompose multi-level parallelism, and also uses a symbolic cost model to search for the optimal parallelism plan.
nnScaler~\cite{nnscaler} offers scheduling abstractions that let experts define flexible search spaces. It leverages existing search strategies to explore the reduced search space and generate parallelism plans, e.g., Alpa for intra-operator parallelism and Tessel~\cite{tessel} for pipeline parallelism.



%% file: content/Limitation.tex
\vspace{-3mm}
\section{Conclusion}

\sysname reduces the complexity of searching parallelism plans for large models by effectively exploiting the structural patterns of large models, thereby enabling an effective profiling-based method for accurate cost estimation and dynamic programming for plan optimization. While it has been proven to be effective in optimizing large model training plans, it has a couple of limitations that should be addressed in future work. First, \sysname does not consider additional levels of parallelism, such as pipeline parallelism. However, as discussed in Sec.\ref{scalability}, its segment sub-search space building and profiling method can be easily extended to pipeline parallel frameworks, providing more accurate stage profiles for pipeline stages partitioning. Second, \sysname does not account for multi-granularity overlapping between communication and computation, which increases uncertainty in downstream optimizations and widens the gap between symbolic cost and actual performance. We believe \sysname remains unaffected as it is based on runtime profiles.

